\pdfoutput=1
\documentclass[11pt,a4paper,twoside,openright,closeany,textany]{book}
\usepackage{makeidx}
\usepackage{graphicx}
\usepackage{paralist}
\usepackage{caption} 
\usepackage{amsmath}
\usepackage{amssymb}
\usepackage{mathtools}
\usepackage{fancyhdr}
\usepackage{multicol}
\usepackage{multirow}

\usepackage{mdframed}
\usepackage[usenames,dvipsnames,svgnames,table]{xcolor}
\usepackage{array}
\usepackage[%
  colorlinks=true,%
  linkcolor=black,%
  urlcolor=black,%
  citecolor=black%
]{hyperref}
\usepackage{nameref} 
\usepackage{ifthen} 
\usepackage[font={small,sl}]{caption}
\usepackage[authoryear]{natbib}
\renewcommand\bibname{References}
\newcommand{\mychapbib}{
  \addcontentsline{toc}{section}{\bibname}
  \bibliographystyle{natbib}
  \bibliography{strucbioinf}
}
\usepackage[colorinlistoftodos]{todonotes}
\usepackage{letltxmacro}

\usepackage{tocstyle} 
\usetocstyle{standard}
\settocfeature{raggedhook}{\raggedright}
\newcommand{\bibfont}{\small\raggedright}
\def\cite{\citep}

\LetLtxMacro{\oldTodo}{\todo}
\renewcommand{\todo}[2][]{\oldTodo[#1]{TODO: #2}}

\usepackage[normalem]{ulem}

\newcommand\inwish[1]{\oldTodo[inline,color=SkyBlue]{WISH: #1}}

\newboolean{onechapter}

\newcommand{\AF}[1][~]{K.\@#1Anton#1Feenstra}
\newcommand{\SA}[1][~]{Sanne#1Abeln}

\newcommand{\HM}[1][~]{Halima#1Mouhib}

\newcommand{\OI}[1][~]{Olga#1Ivanova}

\newcommand{\RB}[1][~]{\mbox{Robbin}#1\mbox{Bouwmeester}}

\newcommand{\JG}[1][~]{\mbox{Jose}#1\mbox{Gavald\'a-Garc\'ia}}

\newcommand{\DG}[1][~]{\mbox{Dea}#1\mbox{Gogishvili}}

\newcommand{\TL}[1][~]{\mbox{Ting}#1\mbox{Liu}}
\newcommand{\IH}[1][~]{\mbox{Isabel}#1\mbox{Houtkamp}}

\newcommand{\orcid}[1]{\href{https://orcid.org/#1}{\raisebox{-0.7ex}{\protect\includegraphics[height=3ex]{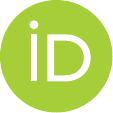}}}}
\definecolor{idgreen}{RGB}{166 206 57}
\newcommand{\mailid}[1]{\href{mailto:#1}{\raisebox{-0.3ex}{\color{idgreen}\textsf{\textbf{\Large \protect@}}}}}

\newcommand{\AFid}{\orcid{0000-0001-6755-9667}}
\newcommand{\SAid}{\orcid{0000-0002-2779-7174}}
\newcommand{\HMid}{\orcid{0000-0001-5031-3468}}
\newcommand{\JGid}{\orcid{0000-0001-6431-3442}}

\newcommand{\OIid}{\orcid{0000-0002-9111-4593}}

\newcommand{\DGid}{\orcid{0000-0001-8809-0861}}

\newcommand{\RBid}{\orcid{0000-0001-6807-7029}}

\newcommand{\TLid}{\orcid{0000-0002-8208-3430}}
\newcommand{\IHid}{\orcid{0000-0002-4222-7292}}

\newcommand{\ACtxt}{Wrote the text}
\newcommand{\ACfig}{Created figures}
\newcommand{\ACref}{Review of current literature}
\newcommand{\ACeds}{Editorial responsibility}
\newcommand{\ACproof}{Critical proofreading}
\newcommand{\ACfb}{Non-expert feedback}

\newcommand{\Angs}[1][~]{\text{\normalfont\AA}}

\renewcommand{\and}{\quad}

\newcommand{\pdbref}[1]{\href{http://www.rcsb.org/pdb/explore.do?structureId=#1}{PDB:#1}}
\newcommand{\arxiv}[2][UNDEFINED]{\href{https://arxiv.org/abs/#2}{\ifthenelse{\equal{#1}{UNDEFINED}}{arxiv.org/abs/#2}{#1}}}

\newcommand{\figref}[2][]{\hyperref[fig:#2]{Figure\@~\ref*{fig:#2}#1}}
\newcommand{\tabref}[1]{\hyperref[tab:#1]{Table \ref*{tab:#1}}}
\renewcommand{\eqref}[2][]{\hyperref[eq:#2]{Equation#1\@~\ref*{eq:#2}}}
\newcommand{\panelref}[2][]{%
    \ifthenelse{\boolean{onechapter}}{%
        \hyperref[panel:#2]{Panel\@~``\nameref{panel:#2}#1''}%
    }{%
        \hyperref[panel:#2]{Panel\@~\ref*{panel:#2}#1}%
    }%
}
\newcommand{\secref}[2][n]{%
    \hyperref[sec:#2]{%
        \ifthenelse{\equal{#1}{n} }{Section\@~\ref*{sec:#2}}{}% just number
        \ifthenelse{\equal{#1}{nn}}{Section\@~\ref*{sec:#2} ``\nameref{sec:#2}''}{}% nm & nr
        \ifthenelse{\equal{#1}{N} }{``\nameref{sec:#2}''}{}% just quoted name
        \ifthenelse{\equal{#1}{NN} }{\nameref{sec:#2}}{}% just name
    }%
}
\newcommand{\chref}[2][n]{%
    \ifthenelse{\boolean{onechapter}}{%
        \ifthenelse{\equal{#2}{ChPref}     }{\arxiv[Chapter ``\nameref*{ch:#2}'']{1801.09442}}{}%
        \ifthenelse{\equal{#2}{ChIntroPS}  }{\arxiv[Chapter ``\nameref*{ch:#2}'']{1801.09442}}{}%
        \ifthenelse{\equal{#2}{ChDetVal}   }{\arxiv[Chapter ``\nameref*{ch:#2}'']{2108.02706}}{}%
        \ifthenelse{\equal{#2}{ChStrucAli} }{\arxiv[Chapter ``\nameref*{ch:#2}'']{1801.09442}}{}%
        \ifthenelse{\equal{#2}{ChDBClass}  }{\arxiv[Chapter ``\nameref*{ch:#2}'']{1801.09442}}{}%
        \ifthenelse{\equal{#2}{ChFunc}     }{\arxiv[Chapter ``\nameref*{ch:#2}'']{1801.09442}}{}%
        \ifthenelse{\equal{#2}{ChIntroPred}}{\arxiv[Chapter ``\nameref*{ch:#2}'']{1712.00407}}{}%
        \ifthenelse{\equal{#2}{ChHomMod}   }{\arxiv[Chapter ``\nameref*{ch:#2}'']{1712.00425}}{}%
        \ifthenelse{\equal{#2}{ChSSPred}   }{\arxiv[Chapter ``\nameref*{ch:#2}'']{1801.09442}}{}%
        \ifthenelse{\equal{#2}{ChFuncPred} }{\arxiv[Chapter ``\nameref*{ch:#2}'']{1801.09442}}{}%
        \ifthenelse{\equal{#2}{ChIntroDyn} }{\arxiv[Chapter ``\nameref*{ch:#2}'']{1801.09442}}{}%
        \ifthenelse{\equal{#2}{ChThermo}   }{\arxiv[Chapter ``\nameref*{ch:#2}'']{1801.09442}}{}%
        \ifthenelse{\equal{#2}{ChMD}       }{\arxiv[Chapter ``\nameref*{ch:#2}'']{1801.09442}}{}%
        \ifthenelse{\equal{#2}{ChMC}       }{\arxiv[Chapter ``\nameref*{ch:#2}'']{1801.09442}}{}%
    }{
    \hyperref[ch:#2]{%
        \ifthenelse{\equal{#1}{n} }{Chapter \ref*{ch:#2}}{}% just number
        \ifthenelse{\equal{#1}{nn}}{Chapter \ref*{ch:#2} ``\nameref{ch:#2}''}{}% name & number
        \ifthenelse{\equal{#1}{N} }{``\nameref{ch:#2}''}{}% just name
      }%
  }%
}
\newcommand{\chrefname}[1]{\hyperref[ch:#1]{Chapter \ref*{ch:#1} ``\nameref{ch:#1}''}}
\newcommand{\partref}[1]{\hyperref[#1]{Part \ref*{#1}}}
\newcommand{\appref}[1]{\hyperref[app:#1]{Appendix \ref*{app:#1}}}

\newcommand{\figsource}[1]{\protect\footnote{Figure source location: \url{#1}}}

\renewcommand{\arraystretch}{1.3}

\makeatletter \@beginparpenalty=5000 \makeatother

\newenvironment{bgreading}[1][]{
  \begin{mdframed}[%
      outerlinewidth=0,%
      linecolor=CornflowerBlue!30,%
      backgroundcolor=CornflowerBlue!30,%
      innerleftmargin=14,%
      innerrightmargin=14,%
    ]
	\ifthenelse{\equal{#1}{}}{}{
        \stepcounter{panel}
    	\subsection*{#1} 
    }
}{%
  \end{mdframed}
}

\usepackage{listings}
\definecolor{backcolour}{rgb}{0.95,0.95,0.92}
\definecolor{codegreen}{rgb}{0,0.6,0}
\definecolor{codegray}{rgb}{0.5,0.5,0.5}
\definecolor{codered}{rgb}{0.8,0,0.0}
\definecolor{codeblue}{rgb}{0.0,0,0.8}
\lstdefinestyle{codeStyle}{
    backgroundcolor=\color{backcolour},   
    commentstyle=\color{codegreen},
    keywordstyle=\color{codeblue},
    numberstyle=\tiny\color{codegray},
    stringstyle=\color{codegray},
    numbers=left,                    
    tabsize=2
} 
\makeindex

\begin{document}

\setboolean{onechapter}{true}

\pagestyle{fancy}
\lhead[\small\thepage]{\small\sf\nouppercase\rightmark}
\rhead[\small\sf\nouppercase\leftmark]{\small\thepage}
\newcommand{\innerfoot}{\footnotesize{\sf{\copyright} Feenstra \& Abeln}, 2014-2023}
\newcommand{\outerfoot}{\footnotesize \sf Intro Prot Struc Bioinf}
\lfoot[\outerfoot]{\innerfoot}
\cfoot{}
\rfoot[\innerfoot]{\outerfoot}
\renewcommand{\footrulewidth}{\headrulewidth}

\mainmatter
\setcounter{chapter}{2}
\chapterauthor{\OI~\OIid \and \JG~\JGid \and \DG~\DGid \and \IH~\IHid \and \RB~\RBid \and 
 \AF*~\AFid~~~~\SA*~\SAid}
\chapterfigure{\includegraphics[width=0.9\linewidth]{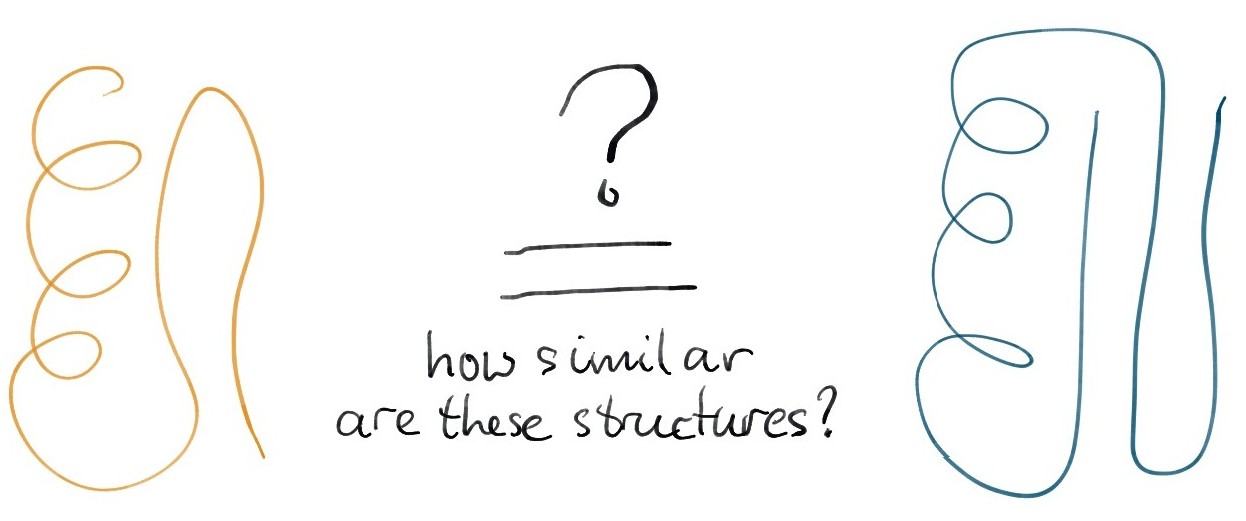}}
\chapterfootnote{* editorial responsability}
\chapter{Structure Alignment}
\label{ch:ChStrucAli}

\ifthenelse{\boolean{onechapter}}{\tableofcontents\newpage}{}

\section{Comparing protein structures}
The Protein DataBank (PDB) contains a wealth of structural information \cite{Berman2000}.  In order to investigate the similarity between different proteins in this database, one can compare the primary sequence through pairwise alignment and calculate the sequence identity (or similarity) over the two sequences. This strategy will work particularly well if the proteins you want to compare are close homologs.  Knowledge of sequence similarity is widely used in for example the Pfam database (\citealt{Finn2014}, see \chref{ChDBClass}): Pfam cluster similar proteins into families based on the sequence of the protein. However, in this chapter we will explain that a structural comparison through structural alignment will give you much more valuable information, that allows you to investigate similarities between proteins that cannot be discovered by comparing the sequences alone.

\begin{figure}[hb]
\centerline{
\includegraphics[width=\linewidth]{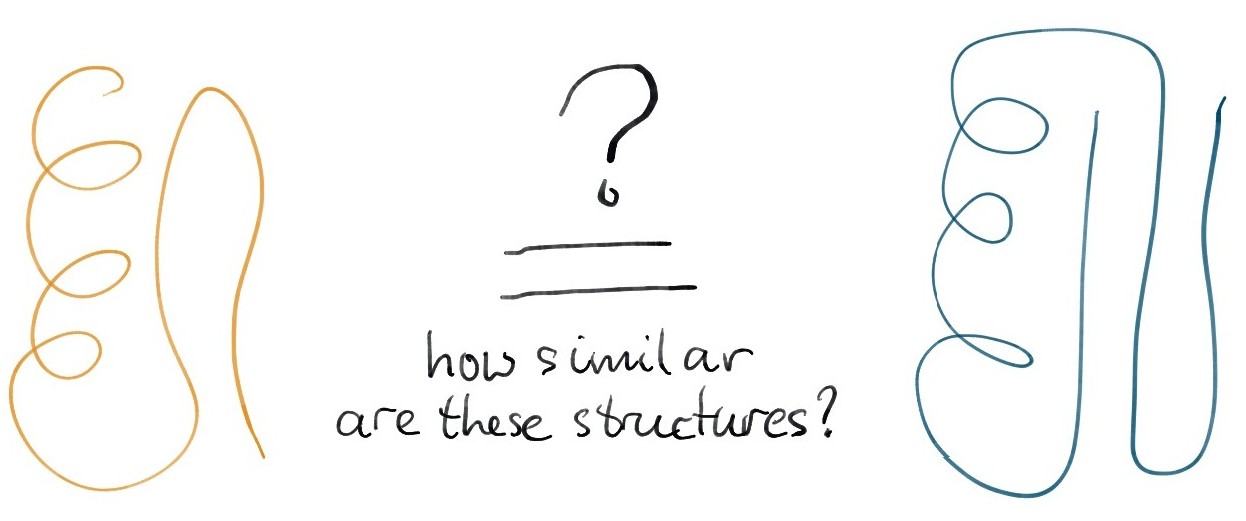}
}
\caption{Structural alignment deals with the problem of determining how similar two structures are -- based on the atomic coordinates alone (no sequence information). }
\label{fig:ChStrucAli-ComparingTwo}
\end{figure}

Furthermore, we will discuss the challenges in understanding how similar two structures are based on \textbf{structural information alone} (i.e.\@ the atomic coordinates, see also \figref{ChStrucAli-ComparingTwo}); this means we do not use any sequence information on the two proteins. Using solely the structure for comparison poses a difficult computational problem due to the many possible ways to align the structure. However, structural comparison is generally considered to be more reliable and evolutionary accurate than a comparison based on sequence similarity.

\subsection{Structure is more conserved than sequence}

One of the reasons why structural comparison is valuable, is that structure is generally  more conserved than sequence. In evolution, selection pressure acts on the function of a protein. If unfolded, the majority of proteins will lose their function. Moreover, it may be dangerous for the cell if a protein is partially folded or misfolded because hydrophobic residues would be exposed leading to the protein aggregation. This effectively puts an evolutionary pressure on the ability of a protein to fold into a functional state and on the folding process itself \cite{Tartaglia2009}.

Structure being more conserved than sequence has several important consequences for methods in structural bioinformatics. This importance is for example illustrated by the large number of structure prediction methods that use these principles as a foundation, and also by the design principles of structural classification databases. Importantly, distant evolutionary relations between two proteins are more easily observed by comparing structures rather than sequences.

\begin{figure}
\centerline{\includegraphics[width=\linewidth]{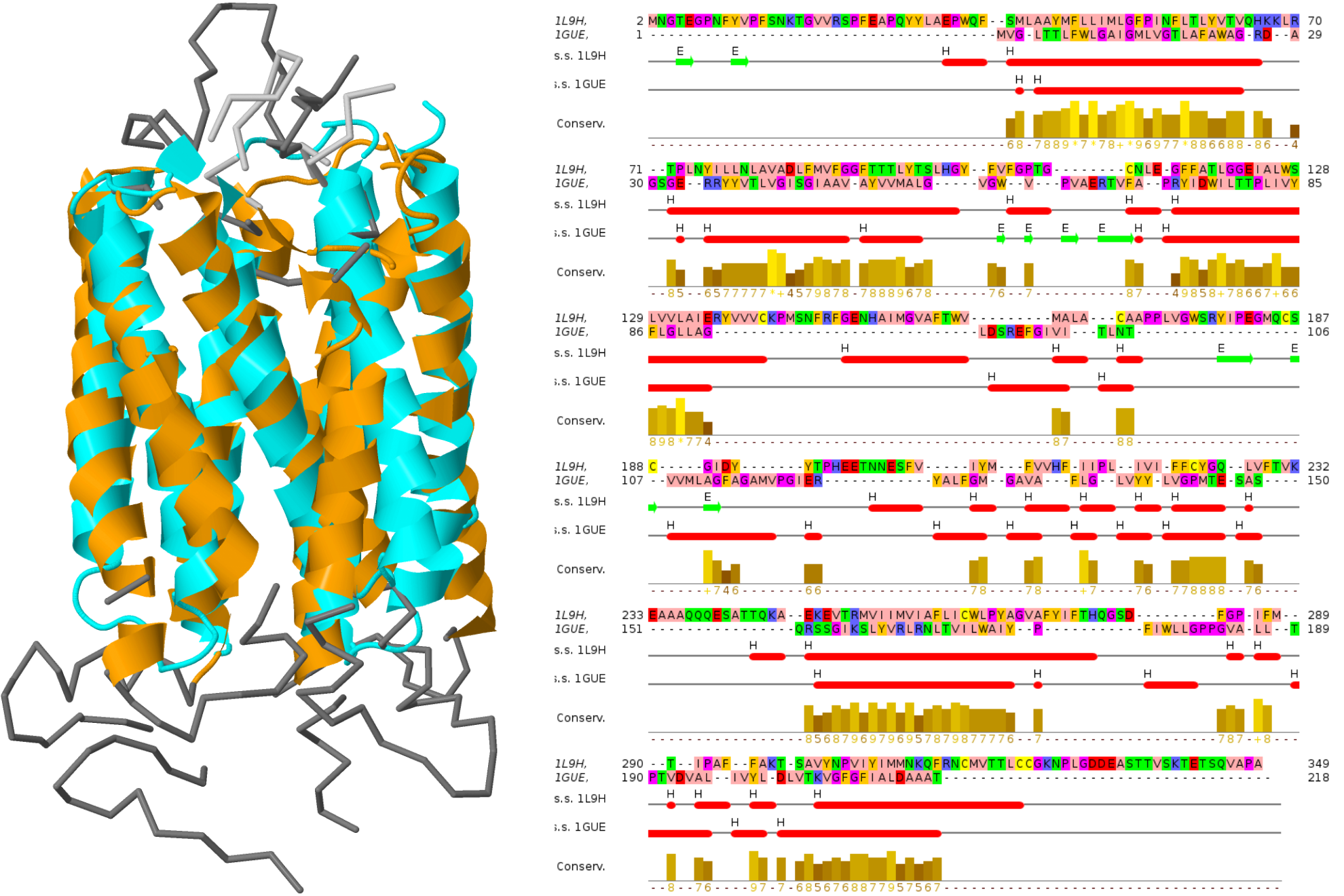}}
\caption{Structure is more conserved than sequence. Left: The output of a structural alignment program, Combinatorial Extension (CE). These two proteins (orange and cyan) have, as you see, a similar structure. They are both rhodopsins, and they have a similar function (light detection). However,their sequence identity (right) is \textbf{less than 5\%}. 
This is below the similarity you would expect from two random sequences. Note that one would not be able to align these proteins using sequence identity alone. One can see that the positioning of the helices are very well conserved between the two structures, but that there is much more variability in the loops (both in structure and in length). The two proteins are bovine rhodopsin (\pdbref{1L9H}, in orange) and sensory rhodopsin (\pdbref{1GUE}, in cyan).
Website at \url{http://www.rcsb.org/pdb/workbench/workbench.do}
}
\label{fig:ChStrucAli-OuptutCE}
\end{figure}

An example of structure conservation without sequence conservation is demonstrated in \figref{ChStrucAli-OuptutCE}, which shows the typical output of a structural alignment program. Two homologous proteins, i.e.,\@ with common ancestor, are aligned with the structural alignment program Combinatorial Extension \cite[CE, by][on the left]{Shindyalov1998} and superimposed on top of each other (on the right). Note that the sequence identity between the proteins is very low; so low in fact, that two \emph{random} sequences aligned by sequence similarity would give a similar score. However, from the structural comparison in \figref{ChStrucAli-OuptutCE}, it is clear that the structures are much alike. In this case, only the structural alignment allows the correct identification of common ancestry for these two proteins.

\section{Structural superposition}

\begin{figure}
\includegraphics[width = \linewidth]{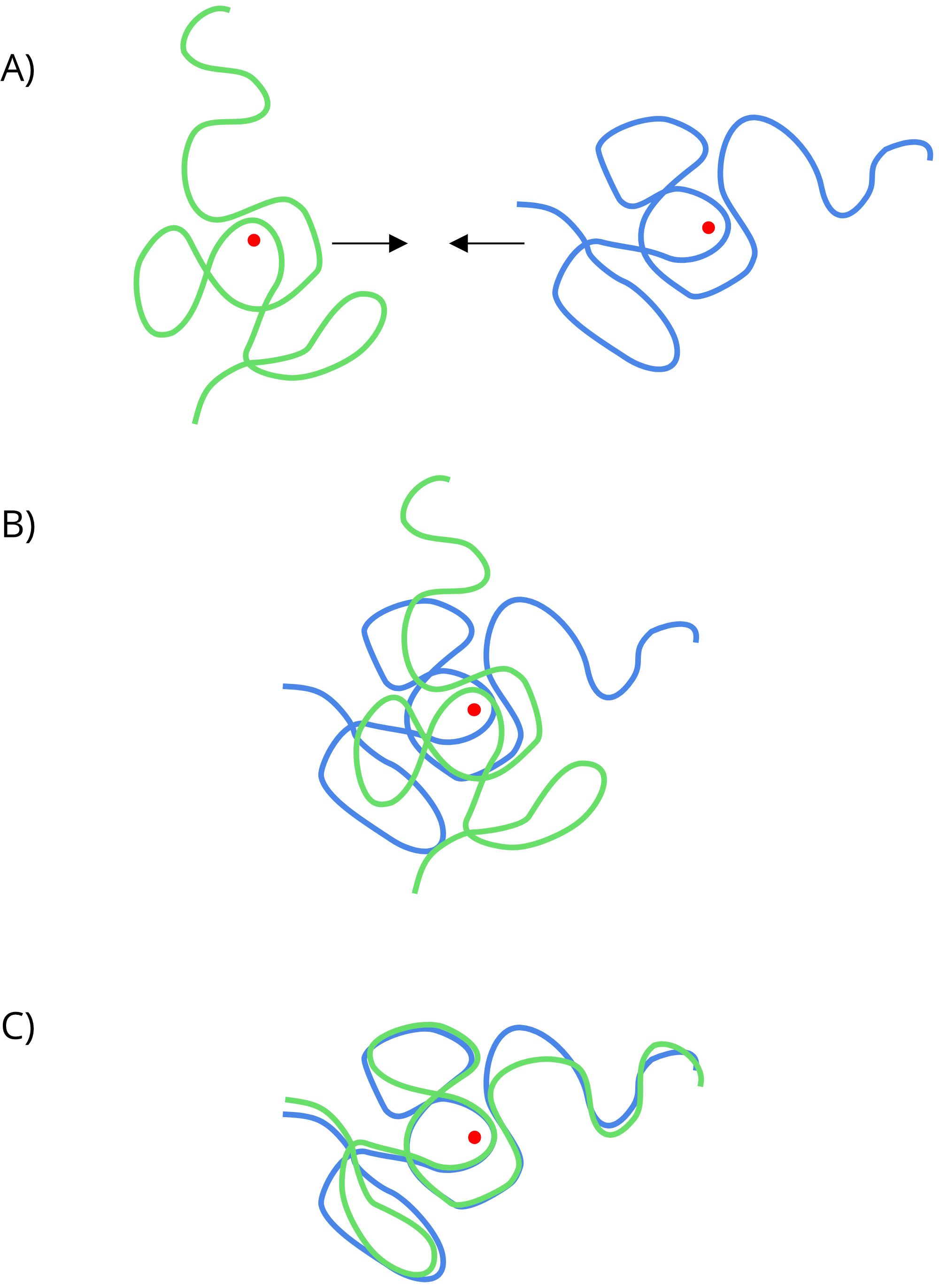}
\caption{\textbf{Superimposing two protein structures.} The lines represent proteins that need to be aligned and the red dots indicate their calculated center of mass. The superposition problem is explained by figures A), where we need to find the "best" overlay in which the the two structures can be compared. C) shows a solution to the problem.
The process of superimposing two structures: A) The centers of mass for the proteins are calculated using \eqref{centre_of_mass}. B) The centers of mass of both proteins are put in the same spatial coordinate. C) The protein structures are superimposed.}
\label{fig:ChStrucAli-center_align}
\end{figure}

The simplest manner to compare two different configurations of a protein is by generating a  \emph{structural superposition} of the two structures. In this section we will focus on the superposition problem, i.e. how to "overlay" two protein structures in 3D space (see \figref{ChStrucAli-center_align}). Keep in mind that if we want to superimpose two structures, we need a mapping between the residues of each structure ( i.e.\@ an alignment - but not based on the sequence). In the next section, \emph{Structural alignment}, we will see how we can compare structures if we do not have an alignment.

\subsection{PDB coordinates as a structure representation}
In \chref{ChDBClass} we will see how a protein structure may be represented and stored in a PDB record in more detail. For now it is enough to know that the structure is recorded in the PDB file by assigning an x, y and z coordinate for each atom (\eqref{rmsd}). Hence each atom can be represented by a positional vector ${\bf p} \in \mathbb{R}^3$ in three dimensional space, such that ${\bf p}=(p_x,p_y,p_z)$. 

Note that a protein structure can be rotated as a rigid body, without changing the interatomic distances. It is up to the experimentalist who created the PDB file to record what the frame of reference is, as any arbitrary rotation and translation may be given. Comparing the rotation and translation of two different protein structures is a key part of the structural superpositioning. Before we go into any more detail, we will first need a score to evaluate the similarity of two protein structures.

\subsection{A score for comparing protein structures -- RMSD}
The root mean square deviation (RMSD), is one of the simplest measures to score the similarity of two protein structures. RMSD calculates the squared difference between two sets of atoms.  In practice, a single representative atom per residue is chosen, i.e., C-alpha or C-beta atoms.
The RMSD of two structures $V$ and $W$, given the residues mapping $M$, is given by: 
\begin{equation}
\begin{split}
\text{RMSD}({V}, {W}) &= \sqrt{ \frac{1}{n} \sum_{i \in M} \|{\bf v_i} - {\bf w_i}\|^2 } \\
    &= \sqrt{ \frac{1}{n} \sum_{i \in M} (v_{i,x} - w_{i,x})^2 + (v_{i,y} - w_{i,y})^2 + (v_{i,z} - w_{i,z})^2}
\end{split}
\label{eq:rmsd}
\end{equation}
Here we calculate the squared distance between atoms ${\bf v_i}$ of structure $V$ and atoms ${\bf w_i}$ of structure $W$. Note the sum over $i$ runs over all matched pairs $M$ of atoms in the alignment of $V$ and $W$; the total number of matched (aligned) residues is $n$. This means we need to know how the residues correspond in the two different structures in order to calculate this score, i.e.\ we need to know which $w_i$ we need to subtract from which $v_i$.

\subsection{Structural superposition and RMSD}

The method of \textbf{finding an optimal rotation and translation} is called \textbf{structural superposition}. An RMSD score between the two sets of atoms is minimised to determine the optimal rotation. Hence, a superposition algorithm provides the rotation that yields the best RMSD fit. Importantly, structural superposition cannot identify the alignment or residues between two protein structures; we will come back to this at the end of this section. 

Note that the RMSD score only gives a measure of the dissimilarity between two protein structures, if they have first been superimposed; without superposition, one would merely measure how far apart the structures are in the given coordinate frame(s), as explained in \figref{ChStrucAli-center_align}. More formally, a measure for structural dissimilarity should be described as the `RMSD after least-squares-fitting', instead of simply using `RMSD', but this is typically omitted in research papers (and even most text books).

\begin{figure}
\includegraphics[width = \linewidth]{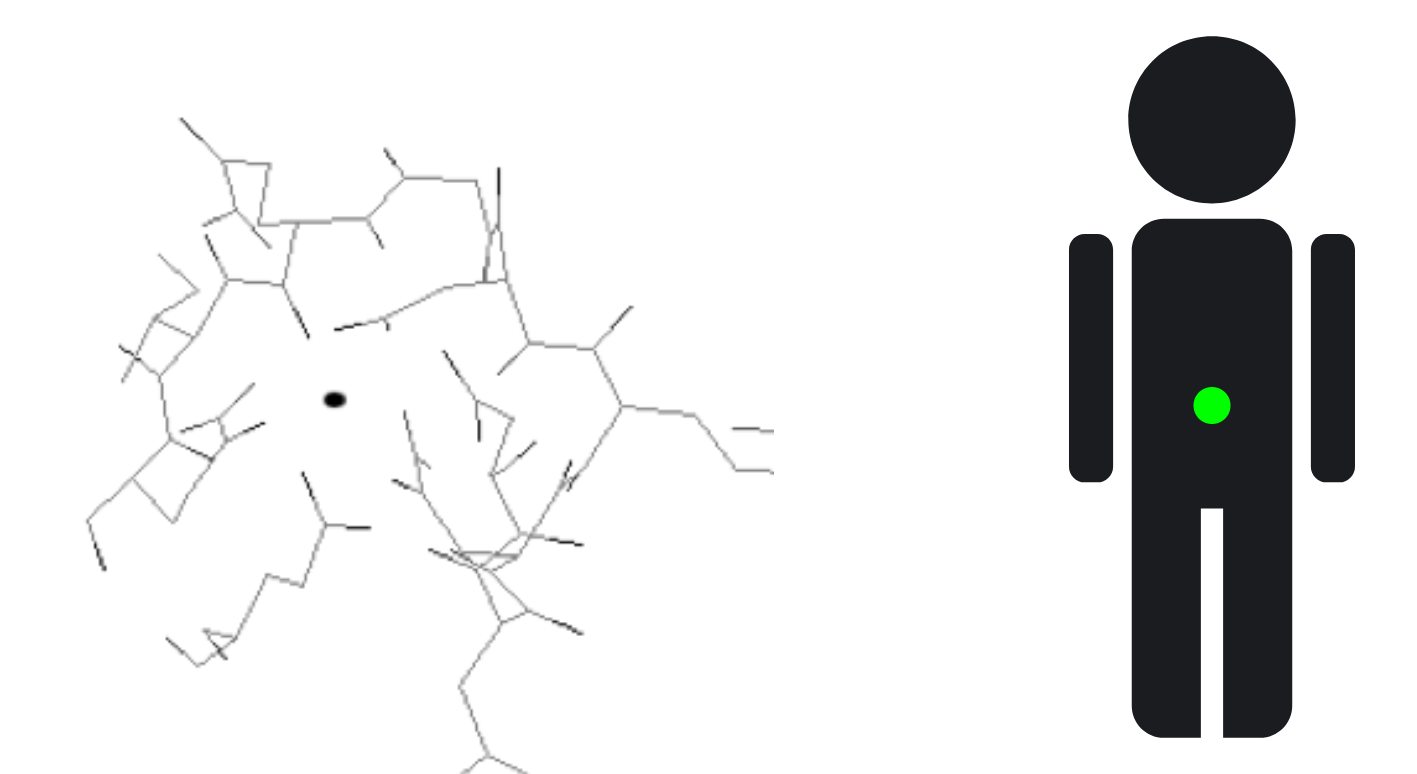}
\caption{\textbf{Center of mass in molecules and humans.} The center of mass is the spatial position determined by the average of all the atoms in a system. The center of mass of a molecule (left, black dot) is calculated the same way that a center of mass for a human (right, green dot) would be calculated. Without going into further detail, the center of mass is placed in the same location as the center of gravity for most objects on Earth, including our bodies.}
\label{fig:ChStrucAli-center-of-mass}
\end{figure}

\figref{ChStrucAli-center_align} also illustrates how one of the two structures needs to be translated and rotated in order to obtain a minimal RMSD between corresponding atom sets. The optimal translation can be found by simply ensuring that the two centers of mass will fall on top of each other. The coordinates for the center of mass of each structure, ${\bf R} \in \mathbb{R}^3$, may be found using \eqref{centre_of_mass}.
\begin{equation}
{\bf R} = \dfrac{\sum_{i=0}^{i=N} m_{i} {\bf r_i}}{\sum_{i=0}^{i=N} m_{i}}
\label{eq:centre_of_mass}
\end{equation}
This is the mass-weighted average of the atomic positions ${\bf r_i}$; here $N$ is the total number of atoms, and $m_i$ is the mass of each atom $i$. \figref{ChStrucAli-center-of-mass} exemplifies the meaning of the center of mass. 

Since we are typically using a single type of atoms (i.e.\@ C-alpha), the masses ($m_i$) in \eqref{centre_of_mass} cancel out. Now we can calculate the directional difference between the two centers of mass to calculate the required translation. Note that a translation in three-dimensional space means that one needs to add or subtract an identical vector, in this case corresponding to the center of mass, from the coordinates of all atoms.

Once the centers of masses have been superimposed, there exists a single rotation that will minimize the RMSD. This rotation can be found exactly; in practice, you need to transform three-dimensional coordinates, to 4D quaternions \cite{Kearsley1989}, and solve an eigenvalue problem using the Jacobi algorithm. We will not go through this method in detail, as it provides little extra understanding of the practical problem. Computationally, it is not very expensive to solve this problem: the exact solution can be found in polynomial time.

Structural super-positioning can be used to compare protein structures for which the mapping of residues is already known. For example, if we want to compare snapshots in a simulation to the native protein structure, we want to actually compare a protein with itself; hence the mapping of the residues is trivial. Similarly, if we want to compare a model from a structure prediction method, with the experimentally determined structure, we would also use structural super-positioning directly. However, if we want to compare (potentially) homologous proteins, the mapping of residues may not be known, and we need structural alignment to generate such a mapping.

\begin{figure}[th]
\includegraphics[width = \linewidth]{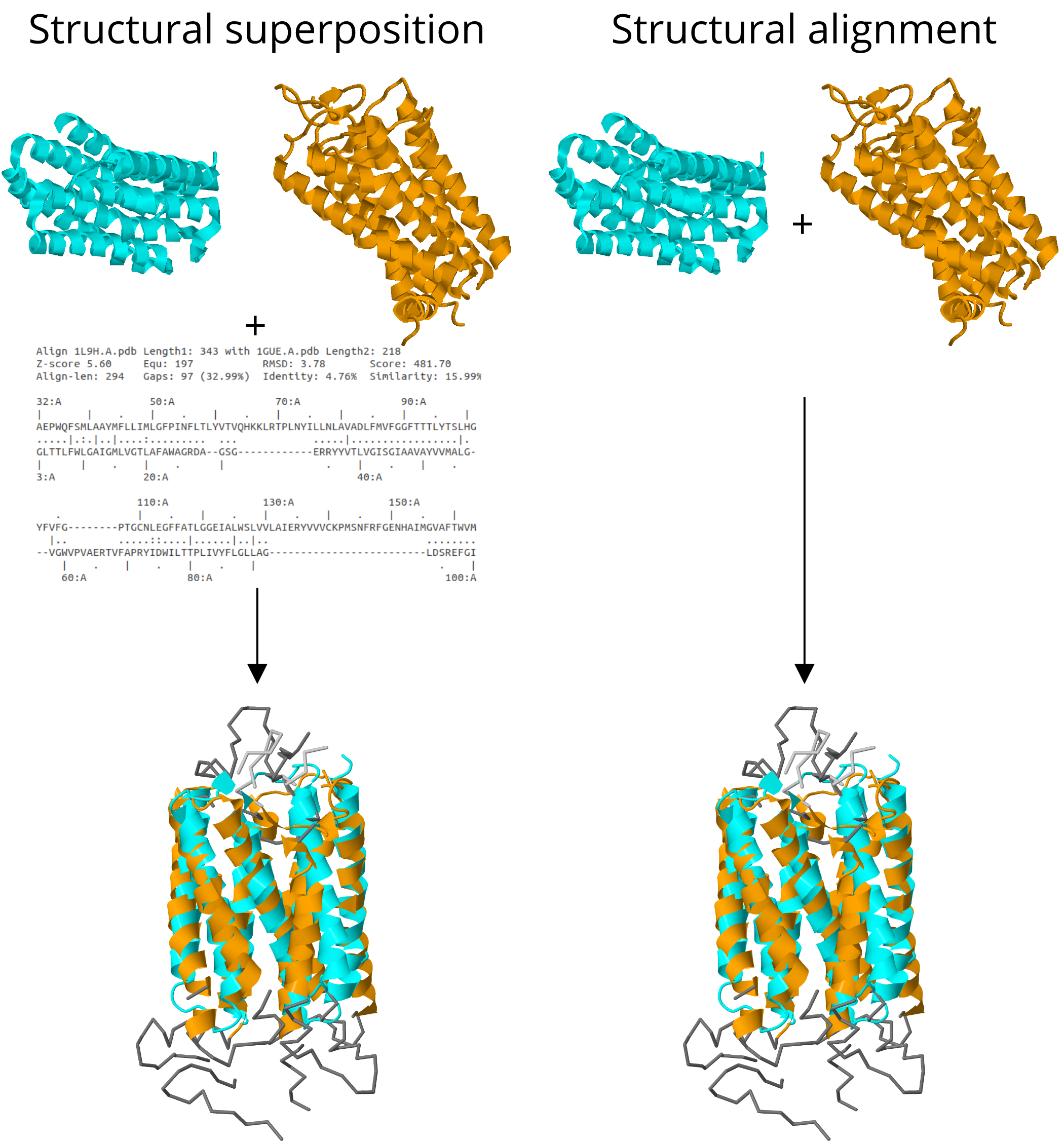}
\caption{\textbf{Structural superposition versus structural alignment.} Left: Structural superposition requires the structures of the proteins and an alignment of the residues as input. Note that if the two structures originate from one protein (and thus have the same sequence) the alignment of the residues is trivial. The superposition method works by minimizing the RMSD, for which we need a mapping (alignment) between the residues. The structural superposition will return two structures in the same frame of reference, such that the RMSD may be calculated. 
Right: Structural alignment takes the protein structures as its only input. The method will try to match similar substructures between the proteins. It will return an alignment, as well as a score for the (dis)similarity of two protein structures. }
\label{fig:align_superimp}
\end{figure}

\section{Structural Alignment}

Structural alignment deals with the problem of generating a mapping between residues, or an alignment, based solely on the structure of two proteins. Note that this mapping of residues is not a trivial task. One may be tempted here to use the sequence as a reference. However, as illustrated by \figref{ChStrucAli-OuptutCE}, this would be a poor choice, as the sequence similarity may be very low for similar structures. Hence we would like to use structure alone to map the corresponding residues (e.g.\@  C-alpha or C-beta atoms) of two structures to be superimposed. The problem of \textbf{finding the optimal match or alignment between the residues of two protein structures} is called \textbf{structural alignment}. 

A typical output of a structural alignment method will therefore provide an alignment of the residues, based on the structure of the backbone alone. Typically, also a superposition of the two structures will be given as an output, as well as a structural alignment score. 
Please also consider \figref{align_superimp} for an explanation of the difference between structural alignment and structural superposition.

The problem of structural alignment is computationally much more difficult to solve than the superposition problem. In fact, it has been suggested to belong to a class of computational problems that is called NP-hard \cite{Hasegawa2009}. In practice, this means that the time required for the computation grows exponentially with the size of the input (protein lengths in this case). A consequence is that for many real sized proteins, of about 200-300 residues, it would be very difficult to find the optimal solution, since it is computationally too expensive to search through all possible structural alignments. However, some methods can provide exact solutions for real size proteins \cite{Wohlers2012}.

\begin{bgreading}[Computational complexity of structural alignment versus sequence alignment]
Note that the pairwise structural alignment problem is very different in terms of computational complexity from the pairwise sequence alignment problem. For the latter problem we can find an optimal solution, by filling in the dynamic programming matrix. Using dynamic programming, the pairwise sequence alignment problem can be solved exactly in $O(n \cdot m)$ time for aligning one sequence of length $n$ with one of length $m$, since $n \times m$ operations need to be calculated to fill in the dynamic programming matrix.  

The difference between these two alignment problems is that there are no natural local bounds in structural alignment. Imagine that we have an already aligned region of the sequence. For sequence alignment, if we change the alignment outside this region, no change of scoring will take place inside the already aligned region. However, with structural alignment, the introduction of a gap outside the aligned region may also affect the score of the already aligned region, due to residues outside the already aligned region that are close in 3D space (but not in sequence space).
\end{bgreading}

\subsection{The three key components of structural alignment}
Structural alignment methods all contain three crucial parts: a suitable \emph{structural representation} for the proteins, a method to \emph{optimize} a similarity measure and a (statistical) \emph{score for protein structure similarity} \cite{Marti-Renom2009}. Plenty of different representations, methods and measures have been used to tackle the problem of structural alignment; here, we will give a brief overview of some of the most important strategies.

\subsection{Structure representation and contact maps}
In order to find a good structural alignment, methods typically encode the structure of both proteins into a representation that can be more easily compared between residues of the two proteins. Note that it is important that this representation is invariant to the frame of reference, otherwise it difficult to compare substructures; hence, the c-alpha coordinates are not suitable. 

An example of a suitable representation is the \emph{contact map}, or contact matrix. Note that such a contact map is defined for a single protein structure, not for a pair of structures. 

\begin{figure}
\begin{center}
\includegraphics[width = \linewidth]{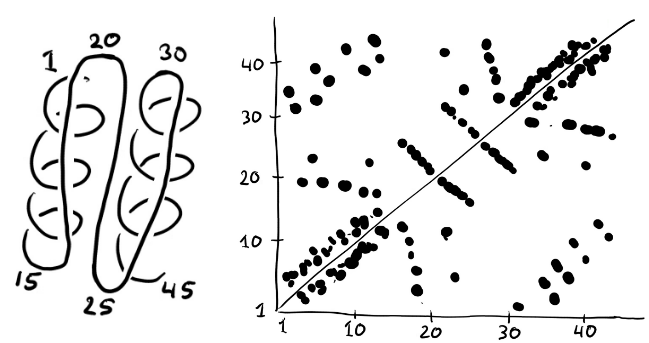}
\end{center}
\caption{\textbf{Contact map of a single protein.} Here it can be observed which atoms are ``in contact'' (closer than a set distance) in a protein structure. From this figure it can be easily observed the the alpha helices close to the main diagonal, and the interaction between residues due to the tertiary structure.}
\label{fig:ChStrucAli-ContactMap}
\end{figure}

A contact can be defined as two residues that are close in three-dimensional space. We can now create a contact map, by considering for each pair of residues in a protein if they make a contact, see for example \figref{ChStrucAli-ContactMap}. 

A contact matrix, $C_{i,j}$, can be mathematically formalized as follows.

\begin{equation}   
C_{i, j} = \begin{dcases*}   1 & \text{if i and j make contact}\\   0 & \text{otherwise}\\   
\end{dcases*}
\label{eq:contactAli}
\end{equation}
Here all possible residues pairs ($i$,$j$) are considered in a single protein structure. A cutoff of around 7 \AA  is typically used to define which residues are in contact.

From such a map, secondary structure elements may be recognized: for example, alpha-helices can be identified as a diagonal line adjacent to the diagonal. This is because in a helix there is a typical hydrogen bonding pattern of residue $i$ and $i+4$ along the sequence, see also \figref{ChStrucAli-ContactMap}.

 We can compare the local three-dimensional surroundings of a residue by considering the contacts. If residues in two different protein structures have (locally) similar contact maps, they have similar substructures. Using contact maps, the structural alignment problem can be redefined as the problem to find an optimal alignment between two contact maps.

Similarly, a \emph{distance map} or distance matrix may be defined for a protein structure. In this case the matrix elements represent the distance between residues, instead of the boolean value indicating if there is a contact or not.  The structural alignment method Dali \cite{Holm2016}, which is still considered to be highly accurate to this day, uses distance matrices to represent structures
Other examples of structure representation are c-beta vectors as used in SSAP\cite{Taylor1989, Orengo1996} and URMSD vectors as used by MAMMOTH \cite{Lupyan2005}.

\subsection{Heuristic optimization algorithms} 

Once we have a good structural representation in which two structures, or the substructures thereof, can be compared, we can formulate structural alignment as an optimisation problem and search for its solution. In the case of the contact matrix, we would look for an alignment of residues in which the corresponding residues have the most similar contact patterns. Since, searching for this optimal solution is thought to be an NP-hard problem (see previous section), performing an exhaustive search for the optimization problem is computationally not feasible. Instead, one needs to use a heuristic search strategy. Heursitic methods search for a good alignment, but cannot guarantee that the solution is optimal. Many of such heuristic search algorithms exist.  Examples used for structural alignment are the branch and bound algorithm \cite{Wohlers2012} and Monte Carlo optimization. In addition, double dynamic programming \cite{Orengo1996} and (single) dynamic programming \cite{Kempen2023} are also used as optimisation strategies. 

\subsection{Statistical scoring of structural alignments}

Next to the actual alignment, a structural alignment method will also provide a (dis)similarity measure for the two structures as an output. This is typically based on the alignment score, and is closely related to the used structure representation. 

We described previously how RMSD can be used to score a structural superposition. Once a structural alignment is made, we could use RMSD as a measure to indicate how similar two proteins are.  Most structural alignment programs will provide the RMSD as an output value for the superposition of the best structural alignment found. 

However, there are some caveats with using RMSD as a measure for structural similarity. Note that, RMSD between two protein structures depends on the size of the proteins being compared. For two random structures, the average distance will become larger if the proteins being compared are large \cite[e.g.][]{Maiorov1995}. A similar effect may occur when one of the proteins that is being compared has many residues far away from the center of mass, for example if it has long loops. Hence, RMSD can be very sensitive to these `outlier' atoms or residues in the structure. 

Apart from the RMSD, structural alignment methods will typically also calculate a p-value or z-score, indicating how significant the structural alignment is. This is important, as the raw comparison scores (e.g.\@ RMSD or overlap in contact maps) are intrinsically length dependent. If we want to see if two structures are significantly similar, this needs to be taken into account: two large structures only sharing two secondary structure elements are typically not homologous.

One can estimate distributions of raw alignment scores over a set of structural alignments between random (non-homologous) structures, for different sizes of structures. From these distributions, subsequent z-scores or p-values may be derived. Note that sequence alignment methods follow a similar strategy to derive statistical scores. For example, BLAST \cite{Altschul1997} uses statistical e-values in which database size, and sequence length are accounted for.

\section {Applications of structure comparison}
Applications of structural comparison and alignment are diverse but mostly concern the fields of evolutionary and structural biology. The majority of structural comparison applications are used to detect remote homology. Entire protein classification databases such as SCOP \cite{Andreeva2008}, CATH \cite{Dawson2017} and also the PDB \cite{Berman2000} use structural alignment methods to find and cluster structurally similar proteins. With increasing database sizes, and the availability of a larger number available predicted structures \cite{Varadi2022}, the speed of these methods becomes increasingly important \cite{Kempen2023}. In \chref{ChDBClass} we will look at some of these resources in more detail.

The identification of distant homology with structure alignment also allows the prediction of protein's functions, \cite{Roy2012} and the classification of proteins with known structures. Structural comparison and alignment ease the prediction of protein structure from the sequence, \cite{Ma2012, Ma2013} and enable the identification of structure patterns and the observation of protein folding space \cite{Yang2000, Kolodny2006}. A less known but significant application of structural comparison and alignment is the recognition of binding sites \cite{Wass2010}.

\section{Key points}
\begin{compactitem}
\item Structural superposition can be used to calculate the RMSD or another similarity metric between two structures, and requires a mapping between residues to do so. 
\item Structural alignment is used to determine a mapping between residues based on structure alone.
\item Structure provides better insight into evolutionary processes because structure is more conserved than sequence.
\item Contact maps can be used to represent a protein structure and its substructures in a rotation invariant fashion.
\item Heuristic methods for score optimization are employed in structural alignment to make computational costs feasible.
\end{compactitem}

\section{Further reading}
For a full overview please see Structural Alignment Chapter by \citet{Marti-Renom2009} in the book ``Structural Bioinformatics'' by \citet{GuBourne}.

\section*{Author contributions}
{\renewcommand{\arraystretch}{1}
\begin{tabular}{@{}ll}
\ACtxt: &   OI, DG, SA, KAF \\
\ACfig: &   JG, SA, KAF \\
\ACref: &   OI, RB, SA \\
\ACproof:&  BS, SA, IH \\
\ACfb:  &   HM, TL, RB \\
\ACeds: &   KAF, SA
\end{tabular}}

\noindent
The authors thank \HM~\HMid{}, \RB~\RB, and \TL~\TLid{} for critical proofreading.

\mychapbib

\clearpage

\cleardoublepage

\end{document}